\documentstyle[twocolumn,prd,aps]{revtex}
\newcommand{\beq}{\begin{equation}}
\newcommand{\eeq}{\end{equation}}
\newcommand{\tr}{{\rm tr}}
\newcommand{\bt}{\widetilde{b}_6}

\def\ibid#1#2#3{{\it ibid.} {\bf #1}, #2 (#3)}
\def\npb#1#2#3{Nucl. Phys. B {\bf #1}, #2 (#3)}
\def\npsb#1#2#3{Nucl. Phys. Proc. Suppl. B {\bf #1}, #2 (#3)}
\def\plb#1#2#3{Phys. Lett. B {\bf #1}, #2 (#3)}

\def\prd#1#2#3{Phys. Rev. D {\bf #1}, #2 (#3)}
\def\prl#1#2#3{Phys. Rev. Lett. {\bf #1}, #2 (#3)}

\begin{document}
\draft
\title{Two-point functions for $SU(3)$ Polyakov Loops near $T_c$}
\author{Adrian Dumitru and Robert D.\ Pisarski}
\bigskip
\address{
Department of Physics, Brookhaven National Laboratory,
Upton, New York 11973-5000, USA\\
emails: dumitru@quark.phy.bnl.gov and pisarski@quark.phy.bnl.gov\\
}
\date{\today}
\maketitle
\begin{abstract}
We discuss the behavior of two point functions for Polyakov loops in a
$SU(3)$ gauge theory about the critical temperature, $T_c$.  From
a $Z(3)$ model, in mean field theory we obtain a prediction for the
ratio of masses at $T_c$, extracted from correlation functions for
the imaginary and real parts of the Polyakov loop.  This ratio is
$m_i/m_r = 3$ if the potential only includes terms up to quartic
order in the Polyakov loop; its value changes as pentic and hexatic 
interctions become important.  The Polyakov Loop Model then predicts how 
$m_i/m_r$ changes above $T_c$.
\end{abstract}

Consider the behavior of an $SU(3)$ gauge theory, without
dynamical quarks, at nonzero temperature.  
The usual quantity measured by the Lattice is the
pressure.  While this is all one needs for thermodynamics in equilibrium,
near equilibrium how correlation functions behave is also of importance.

In this note we discuss a basic, gauge-invariant
correlation function, the
two-point function of Polyakov loops.  We work with a $Z(3)$
model \cite{svetitsky,cargese}; near the critical temperature,
this is equivalent to the Polyakov Loop 
Model~\cite{plm,eloss,wirstam,proceedings}.

At a nonzero temperature $T$, the thermal Wilson line is:
\begin{equation}
{\bf L}(\vec{x}) \; = \; 
{\cal P} \exp \left( i g \int^{1/T}_0 A_0(\vec{x},\tau) \, 
d\tau \right) \; ,
\end{equation}
with $A_0$ in the fundamental representation.
This transforms as an adjoint field under the local $SU(3)/Z(3)$ 
gauge symmetry, and as a field with charge one under the global
$Z(3)$ symmetry.  To obtain a gauge invariant operator, the
simplest thing to do is to take the trace, forming the Polyakov loop,
\begin{equation}
\ell \; = \; \frac{1}{3} \; {\rm tr}\left( {\bf L} \right) \; .
\end{equation}
This transforms under the global $Z(3)$ symmetry as a field with charge one,
\beq
\ell \rightarrow \exp\left(\frac{2 \pi i}{3}\right) \ell
\label{ea}
\eeq
The expectation value of $\ell$ is only nonzero above $T_c$,
which is the temperature for the deconfining phase transition.

For three (or more) colors, $\ell$ is a complex valued field.
Implicitly, we assume that the expectation value of $\ell$ is
real.  In this vacuum, we can then ask for the correlation functions
of the real and imaginary parts of the Polyakov loop,
\beq
\ell_r \; = \; {\rm Re}\; \ell \;\;\; , \;\;\;
\ell_i \; = \; {\rm Im}\; \ell \; ,
\eeq
which do not mix.
As the theory is globally $Z(3)$ invariant, we can
define equivalent masses in either of the other two degenerate vacua by
an appropriate overall $Z(3)$ rotation.

We need a standard feature of propagators in coordinate space.
In $d$ space-time dimensions,
at distances large compared to $1/T$, only the $d-1$ spatial dimensions
matter.  For a field of mass $m$, at distances $x\gg 1/m$, of course 
the propagator falls off exponentially,
\beq
\int d^{d-1} p \; \frac{ {\rm e}^{- i \vec{p} \cdot \vec{x} } }{p^2 + m^2} 
\; \sim \; \; \int d^{d-2} p_\perp 
\; {\rm e}^{- x (m +  p_\perp^2/2m)}
\sim \; \frac{{\rm e}^{- m x}}{x^{(d-2)/2}} \; .
\eeq
We are interested not in the familiar exponential fall-off, but in 
power law prefactor of $1/x^{(d-2)/2}$.  While derived in free field
theory, it is valid generally.  It arises from the
fluctuations of a point particle, in the directions perpendicular to
the direction along which we measure it.

In perturbation theory, the Polyakov loop is near unity,
and can be expanded in
powers of $A_0$.  Assuming for the purposes of discussion that
the $A_0$ field is purely static, we obtain
\beq
\ell \; \approx \; 1 \; - \; \frac{g^2}{3 T^2} \; \tr
\left( A_0^2 \right) \; - \; i \; \frac{g^3}{3 T^3} \;
\tr \left( A_0^3 \right) \; + \ldots 
\eeq
Thus the real part of the Polyakov loop starts out as a coupling to 
two $A_0$'s, and the imaginary part of the Polyakov loop, to three
$A_0$'s.  Thus to lowest order in perturbation theory \cite{nadkarni},
\begin{equation}
\langle \ell_{r}(x) \ell_{r}(0) \rangle - \langle \ell \rangle^2 \sim 
\frac{\exp(- 2 m_D x)}{x^2} \; ,
\label{eb}
\end{equation}
\begin{equation}
\langle \ell_{i}(x) \ell_{i}(0) \rangle \sim 
\frac{\exp(- 3 m_D x)}{x^3} \; .
\label{ec}
\end{equation}
Here $m_D$ is the Debye mass; to lowest order in perturbation
theory, $m_D = g T$.   Each result is simply the product of either two
or three $A_0$ propagators over distances $x \gg 1/m_D$.
Thus, {\it if} one ignores the difference in prefactors,
and defines $m_r$ and $m_i$ from the exponential falloff of
(\ref{eb}) and (\ref{ec}), respectively, we conclude that $m_i/m_r = 3/2$.
These estimates could be altered by mixing with the purely magnetic sector
\cite{nonpert}, which we discuss at the end of the paper.

We now turn from the perturbative regime, valid at very high temperatures,
to temperatures near $T_c$.  Instead of dealing with $A_0$,
we construct an effective Lagrangian directly for the
Polyakov loop itself \cite{svetitsky,cargese}.  
The theory must be invariant under the global $Z(3)$ symmetry of (\ref{ea}).  
Implicitly, we assume that the deconfining
transition for three colors is so weakly first order
that it is appropriate
to classify the effective Lagrangian according to a
renormalization group analysis for a second order transition.
At large distances, $x \gg 1/T$, the theory is effectively
three dimensional; by the renormalization group, all relevant
and marginal operators include terms up to 
sixth order in $\ell$.  The most general, renormalizable
Lagrangian in three dimensions, invariant under a 
global $Z(3)$ symmetry, is
$$
{\cal Z}_W \; |\vec{\partial} \ell|^2 
\; - \; \frac{b_2}{2} |\ell|^2 \; - \; \frac{b_3}{3}
\left( \frac{\ell^3 + (\ell^*)^3}{2} \right) 
$$
$$
\; + \; \frac{b_4}{4}\left( |\ell|^2\right)^2 
\; + \; \frac{b_5}{5} \; |\ell|^2 \; 
\left( \frac{\ell^3 + (\ell^*)^3}{2} \right) 
$$
\beq
+ \; \frac{b_6}{6} \left(|\ell|^2\right)^3
+ \; \frac{\bt}{6} \left( \frac{\ell^3 + (\ell^*)^3}{2} \right)^2 \; .
\eeq
The terms which only involve powers of $|\ell|^2$, proportional to
$\sim b_2$, $b_4$ and $b_6$, are each
each invariant under a global symmetry of 
$U(1)$, $\ell \rightarrow \exp(i \theta)\ell$ for
arbitrary $\theta$.  The kinetic term is also $U(1)$ invariant.
For reasons to become clear shortly, 
we include a factor of wave-function renormalization,
${\cal Z}_W$, but suppress 
all other renormalization constants for mass and coupling constants.
The cubic term, $\sim b_3 \ell^3$,
and the pentic term, $\sim b_5 |\ell|^2 \ell^3$, are only 
invariant under the $Z(3)$ symmetry of (\ref{ea}).  
Of the two hexatic interactions, that $\sim b_6$ is $U(1)$ invariant,
while that 
$\sim \bt \ell^6$ is invariant not only under $Z(3)$,
but under a larger symmetry of $Z(6)$, $\ell \rightarrow \exp(\pi i/3) \ell$.  

It is natural to require that the potential is bounded from below for large
values of $\ell$.  At large $\ell$, the hexatic terms
dominate over all other couplings.  For real $\ell$, this gives
\beq
b_6 \; + \; \bt \; > \; 0  \; .
\label{eh}
\eeq
One can also choose $\ell = \exp(i \pi/6) \ell_0$, with real $\ell_0$;
then the term proportional to $\bt$ drops out, giving the condition
\beq
b_6 \; > \; 0 \; .
\label{ei}
\eeq

We perform a mean field analysis, where
all coupling constants are taken as constant with temperature, except
for the mass term, $\sim b_2 |\ell|^2$.  
About the transition, condensation
of $\ell$ is driven by changing the sign of $b_2$
\cite{cargese,plm}.
The sign of $b_3$ is determined by the following.  At very high
temperature, the favored vacuum is perturbative, with $\ell_0 \approx 1$,
times $Z(3)$ rotations.  We then choose $b_3>0$ so that in the $Z(3)$ model,
there is always one vacuum with a real, positive expectation value for
$\ell_0$.

We also choose $b_4 > 0$.  This is not a matter of convention.
With the above potential, 
in mean field theory there are two ways of obtaining a first
order transition.  One is through a non-zero, cubic coupling.  
The other is if the four point coupling is negative.  It is 
natural, however, to assume that $b_3, b_5 \neq 0$ {\it and} $b_4>0$.  
For example, if $b_3 = b_5 = 0$, then 
the unbroken symmetry of the effective Lagrangian is not $Z(3)$, but $Z(6)$.  
While one can then obtain a first order transition by choosing
$b_4$ to be negative, it would also imply
a six-fold degeneracy between vacua in the broken symmetry phase.
Such a degeneracy is not observed, even approximately,
by Lattice simulations in the deconfined phase \cite{lattice}.  Consequently, 
we assume that $b_3, b_5 \neq 0$, and that $b_4$ is positive.
This also agrees with results from the Polyakov Loop Model
\cite{plm,eloss}.

While the value of the pentic coupling constant
is not well fixed, but it cannot 
dominate the hexatic couplings.
If the hexatic interactions vanish, $b_6 = \bt = 0$, then
with $\ell = \exp(i \theta/3) \ell_0$, the pentic term is negative
for $\pi/2 < \theta < 3 \pi/2$, and the potential is unbounded from
below.  Consequently, we consider the pentic interactions as
a small perturbation on the other couplings.

This is an effective Lagrangian in three dimensions, so that
$\ell$ has dimensions of mass$^{1/2}$; then
the coupling constants $b_n$ have dimensions of 
mass$^{(6-n)/2}$.  In particular, $b_6$ and $\bt$ are dimensionless.
To make the equations somewhat simpler, we choose the overall mass
scale so that $b_4 = 1$; expressions for arbitrary values of $b_4$
can be derived from the following by an obvious rescaling.
We remark that in the
Polyakov Loop Model \cite{plm,eloss,wirstam,proceedings}, 
the dimensions of $\ell$, and of the coupling constants, are all
made up by powers of temperature.  Near $T_c$, however, this
doesn't matter.

We now compute the masses for the real and imaginary parts of
the Polyakov loop.
As stated, we choose the vacuum expectation value of $\ell$,
$\langle \ell \rangle = \ell_0$, to be real.  
The equation for $\ell_0$ is just the first derivative of the
potential for $\ell$, 
\beq
-\; b_2 \ell_0 \; - \; b_3 \; \ell_0^2 \; + \; \ell_0^3
\; + \; b_5 \ell_0^4 \; + \; (b_6 + \bt) \ell_0^5 = 0 \; .
\eeq
Computing second derivatives of the potential, the mass squared for
the real part is:
\begin{equation}
m^2_r \; = \; 
- \; b_2 \; - \; 2 b_3 \ell_0 \; + \;
3 \ell_0^2 \; + \; 4 b_5 \ell_0^3 \; + \; 5 (b_6 + \bt) \ell_0^4 \; ,
\end{equation}
while that for the imaginary part is:
\begin{equation}
m^2_i 
\; = \; - \; b_2 \; + \; 2 b_3 \; \ell_0 \; + \; \ell_0^2 
\; - \; \frac{4}{5} b_5 \; \ell_0^3
\; + \; (b_6 - 2 \bt) \ell_0^4 \; .
\end{equation}
The two-point function of Polyakov loops is then, 
\begin{equation}
\langle \ell_{r}(x) \ell_{r}(0) \rangle - \langle \ell \rangle^2 \sim 
\frac{\exp(- m_r x)}{x} \; ,
\label{ed}
\end{equation}
\begin{equation}
\langle \ell_{i}(x) \ell_{i}(0) \rangle \sim 
\frac{\exp(- m_i x)}{x} \; .
\label{ee}
\end{equation}

Notice that unlike (\ref{eb}) and (\ref{ec}), the power law prefactor
is in each case just $\sim 1/x$, from the exchange of a single
$\ell$-field.  Above $T_c$, $m_i \neq m_r$.  

The global $Z(3)$ symmetry is unbroken below $T_c$: $\ell_0 = 0$.
With our conventions, $b_2 < 0$;
the mass squared for $\ell_r$ and $\ell_i$ are positive and equal.
Thus in our $Z(3)$ model, 
below $T_c$ the two-point functions of Polyakov loops
remain as in (\ref{ed}) and (\ref{ee}) at large
distances, with $m_i = m_r$.
We remark that the prefactor of $\sim 1/x$, for
$T < T_c$ and $x \gg 1/T$, 
agrees with calculations in Nambu-Goto string models \cite{forcrand}.
This is true for any number of space-time dimensions:
the transverse motion of a point
particle, and the transverse fluctuations of a world-sheet, both give
the same prefactor of $\sim 1/x^{(d-2)/2}$.
This is an important consistency check on models of Polyakov loops,
such as \cite{plm}, below $T_c$.

When $b_3$ or $b_5$ are nonzero, 
the transition is necessarily of first order.
Henceforth, we restrict ourselves to the critical temperature,
$T_c$.  This occurs when the nontrivial minimum
is degenerate with the trivial minimum; {\it i.e.}, when 
\beq
-\; \frac{b_2}{2} \ell_0^2 
\; - \; \frac{b_3}{3} \; \ell_0^3 \; + \; \frac{1}{4} \; \ell_0^4
\; + \; \frac{b_5}{5} \; \ell_0^5
\; + \; \frac{b_6 + \bt}{6} \; \ell_0^6 = 0 \; .
\eeq
This then becomes an equation which determines the value of $b_2(T_c^+)$.

If we compute each mass individually, $m_r$
and $m_i$, then there is a complication which we have set aside.
The kinetic term for $\ell$ does not have a standard normalization;
at tree level, it equals $3/g^2$, where $g$ is the QCD coupling
constant \cite{plm}.  At one loop order, Wirstam \cite{wirstam} finds that
this wave function renormalization is
\beq
{\cal Z}_W \; = \; \frac{3}{g(T)^2} \left( 1 \; - \; .08 \frac{g^2}{4 \pi} 
\; + \; \ldots \right) \; .
\eeq
Consequently,
the masses measured on the Lattice are actually $1/\sqrt{{\cal Z}_W}$
of the expressions above.  Thus the overall scale of $m_r$ and $m_i$
depend upon the value of $g^2(T)$, which is not easy to extract 
directly from Lattice data.  

To avoid the dependence on ${\cal Z}_W$, instead we compute the ratio
of $m_i/m_r$, in which ${\cal Z}_W$ completely drops out.  It is
this ratio which we think would be especially interesting to compute
on the Lattice.  

These equations can be solved numerically for arbitrary values of
$b_5$, $b_6$, and $\bt$.  It is illustrative to first solve for the case
in which $b_5 = b_6 = \bt = 0$.  For arbitrary $b_2$,
\beq
\ell_0 \; = \; \frac{1}{2} \left( b_3 \; + \; 
\sqrt{4 b_2 \; + \; b_3^2 } \right) \; .
\label{ef}
\eeq
With this $\ell_0$, at the transition temperature, $b_2 = - 2 b_3^2/9$,
and $m_i/m_r = 3$.  This ratio changes as $b_5$, $b_6$, and $\bt$ increase
from zero.  To lowest nontrivial order in these coupling constants,
\beq
\frac{m_i}{m_r} \; \approx \;
3 \; \left( 1 
\; - \; \frac{8}{15} \; b_3 b_5
\; - \; \frac{4}{27} \; b_3^2 
\left( 4 b_6 \; + \; 5 \bt \right) \right) \; + \; \ldots
\eeq
For positive values of
$b_5$ and $\bt$ (remember $b_6$ is necessarily positive), 
the ratio $m_i/m_r$ decreases from
$3$ as $b_5$, $b_6$, and $\bt$ increase.  
It is also easy to obtain $m_i/m_r > 3$, though.
The coupling $b_5$ could be negative; even if $b_5 = 0$,
when $\bt = - b_6$, {\it etc.}

One can also solve for large values of $b_6$ and $\bt$.  
Taking $b_5 =0$, in this case
\beq
\frac{m_i}{m_r} \; \approx \;
\sqrt{\frac{3}{2}}
\; \sqrt{1 + \frac{b_6}{b_6 + \bt}} \; + \ldots \; .
\eeq
The combination of $b_6$ and $\bt$ on the right hand side
is necessarily a ratio of positive numbers in all cases.

Working up from $T_c$, it is also possible to compute $m_i/m_r$
within the context of the Polyakov Loop Model (PLM) \cite{plm}.  In
the PLM, the pressure is determined by the potential of $\ell$
times $T^4$; for constant values of $b_3$, $b_4$, $b_5$, $b_6$, and
$\bt$, the temperature dependence of the pressure then fixes
$b_2(T)$.  Consequently, this also determines how 
$m_i/m_r$ changes with temperature.  This appears to be a crucial
and unambiguous test of the PLM.  For example, if we take
$b_6=\bt=0$, then at very high temperatures, where
$\ell_0 \approx 1$, (\ref{ef}) fixes $b_2 = 1 - b_3$, and~\cite{proceedings}
\beq
\frac{m_i}{m_r} \; = \; \sqrt{\frac{3 b_3}{2 - b_3}} \; .
\eeq
To agree with the perturbative result as $T \rightarrow \infty$, 
$=3/2$, fixes $b_3 = 6/7$.
This is remarkably close to the value found from a fit to the
pressure near $T_c$, which is $b_3 \approx 0.9$ \cite{b3}.  
A more careful analysis,
including the effects of pentic and hexatic interactions, is necessary
in order to make a detailed comparison.

To summarize, naively in the perturbative regime at very
high temperature, $m_i/m_r = 3/2$ \cite{nadkarni}.  
If pentic and hexatic terms in 
$\ell$ are a small perturbation on the quartic terms,
then as $T \rightarrow T_c$, $m_i/m_r$ should increase; 
at $T_c$, how close this ratio is to $3$ then
indicates how important these other coupling constants are.
Conversely, if $m_i/m_r$ does not increase significantly as the temperature
decreases from $2 T_c$ down to $T_c^+$, then while it might be
able to obtain a fit with large values of $b_6$ and $\bt$, to us
this would seem unnatural, and disfavor the present analysis.

Our discussion neglects mixing between the electric sector
and the static, magnetic sector~\cite{nonpert}.
While the former might be perturbative, the latter never is.
Static, magnetic glueballs couple to each
other as a three-dimensional
gauge theory, with coupling constant $g^2 T$.  Hence the
mass of any static, magnetic glueball is some number times
$g^2 T$.  In weak coupling, then, as the Debye mass is $m_D \sim gT$,
electric fields are much heavier than magnetic glueballs.  
The real part of the Polyakov loop, $\sim \tr (A_0^2)$, can then
mix with spin-zero glueballs, such as $\sim \tr(G_{ij}^2)$.  
Also, the imaginary part of the Polyakov loop, 
$\sim \tr (A_0^3)$ can mix
with the spin-zero glueball $\tr(A_0 G^2_{ij})$.
(There is also a spin-one glueball, $\epsilon^{i j k} \tr(A_0 G_{jk})$.
However, the coupling of such a spin-one state to any spin-zero state
must involve one derivative, and so 
vanish at zero momentum.  Spin here 
refers to a classification in three dimensions.)

Simulations from temperatures of $\sim 2 T_c$ on up
have been done by Hart, Laine, and Philipsen \cite{hart};
for related work, see \cite{other}.
They find that the magnetic sector is not lighter than the
electric until temperatures $T > 10^{2} T_c$.
Unexpectedly, they do not see significant mixing between states
such as $\tr (A_0^2)$ and $\tr(G_{ij}^2)$.
For example, at extremely high temperatures
of $10^{11} T_c$, 
the mass of $\tr (A_0^2)$ is approximately equal to that
of $\tr (A_0^3)$; both masses are about $\approx 1.9$ 
times the mass of the
lightest static, magnetic glueball, $\tr(G_{ij}^2)$.
The lightest electric state is the spin-one glueball,
$\epsilon^{i j k} \tr(A_0 G_{jk})$; we note the 
(presumably meaningless) numerical coincidence that its mass is 
$=1.5$ of the lightest static, magnetic glueball to within $0.1\%$,
although the quoted statistical errors are $\approx \pm 6\%$.

At temperatures of $\sim 2 T_c$, the situation is completely reversed:
then the electric sector is {\it lighter} than that for
the static, magnetic sector.  
The lightest state is that for the real part of the Polyakov
loop, $\sim \tr(A_0^2)$.  Relative to this state, the mass
of the spin-zero electric state, $\tr(A_0^3)$, is
$\approx 1.81$; 
that of the spin-one electric state, $\epsilon^{i j k} \tr(A_0 G_{jk})$,
is $\approx 1.76$;
that of the lightest static, magnetic
glueball, $\tr(G_{ij}^2)$, is $\approx 2.4$.

Thus at $2T_c$, the ratio $m_i/m_r \approx 1.81$.  It would be very
interesting to know how $m_i/m_r$ changes at $T: 2T_c \rightarrow T_c$;
we are unaware of data on $m_i$ in this range.  Nevertheless,
at least we can safely neglect mixing of the (light) electric sector
with the (heavy) magnetic sector.

At present, most Lattice data is for the two
point function of complex conjugate Polyakov loops, 
$\langle \ell(x) \ell^*(0) \rangle - |\langle \ell \rangle|^2$.  From 
either the $Z(3)$ model, or perturbation theory, 
in the deconfined phase 
it appears that $m_i$ is always greater than $m_r$.
Thus without extracting excited states, this
correlation function only gives us information on $m_r$, and not
on $m_i$.  

One can also decide whether an effective model of Polyakov loops applies
by seeing if the power law prefactor is $1/x$, 
as in (\ref{ed}), or $1/x^2$, as in (\ref{eb}).  From Lattice
simulations in $3+1$ dimensions by Kaczmarek, Karsch, Laermann, and Lutgemeier
\cite{bielefeld}, the power
law prefactor is very near unity only near $T_c$, and increases
to $\sim 1.4$ by $2 T_c$.  Thus our effective model does appear to
hold in at least some limited region near $T_c$.  
As the exchange of {\it any} massive state generates a prefactor of
$\sim 1/x$, instead of varying the prefactor, to us it seems more
natural to take a sum over several states, each
a prefactor of $\sim 1/x$.  We do admit that this is 
only true in mean field theory, as (nearly) critical
fluctuations introduce nonzero anomalous dimensions, $\eta$,
which change the prefactor to 
$\sim 1/x^{1 - \eta}$.  In three dimensions, however, such
anomalous dimensions $\eta$ are usually very small;
$\eta < .1$ for either $Z(2)$ or $O(2)$ symmetry groups \cite{zinn}.

It is also of interest to consider the transition in $2+1$ dimensions.
While mean field theory suggests a first order transition, fluctuations
in two dimensions are so strong that the transition is in fact of
second order.  Then our mean field analysis is completely inappropriate;
one is at the symmetric phase at $T_c$, with $m_i = m_r$.  
However, correlation functions of both imaginary and real parts of
the Polyakov loop have been computed above $T_c$ by
Bialas, Morel, Petersson, and Petrov \cite{twoplusone}.  
In all cases, the power law
prefactor is found to be $1/\sqrt{x}$, which is what it should be
for single particle exchange in $d=2+1$ dimensions.  Bialas {\it et al.}
have also computed $m_i/m_r$; working down from high temperatures,
this ratio increases to $\approx 2.0$, 
then seems to decrease.  This might arise from
a trade-off between mean field effects, as discussed herein, and
critical fluctuations near $T_c$, special to $2+1$ dimensions.

In passing, we also comment that studies with two colors are also
of interest.  For two colors, $\ell$ is real, and so there is only
$m_r$.  One could test, however, to see whether the relation between
$m_r$ and the pressure, as given by the PLM, holds up \cite{plm}.  
With only one mass to measure, the value of
$m_r$ does depend upon ${\cal Z}_W$, and so $g^2(T)$.  Also,
the effects of critical fluctuations must be incorporated.  On the
other hand, the
potential for $\ell$ only involves two coupling constants, $b_4$, and $b_6$;
$b_3 = b_5 = \bt = 0$.  Still, one basic test is that in $3+1$ dimensions,
the deduced value of $g^2(T)$ increases as $T$ decreases.  Lastly, 
there are speculative reasons for thinking that the width of the critical
region is greater in $2+1$ dimensions 
than in $3+1$ dimensions \cite{cargese}.

We look forward to future measurements of these and other quantities
on the Lattice, near the critical temperature.

{\bf Acknowledgements:}
This research was supported by
DOE grant DE-AC02-98CH10886.
We thank A. D. Hart, J. T.~Lenaghan, S. Ohta, and O. Philipsen
for useful comments and discussions.  We are especially indebted to
M. Laine for pointing out the existence of the pentic coupling.


\begin{references}
\bibitem{b3}
This value of $b_3$ follows from Scavenius, Dumitru, and Lenaghan,
\cite{plm}.  This value follows not only by fitting the potential
for $\ell$ to the pressure, but also requiring that 
$\ell_0 \rightarrow 1$ as $T\rightarrow \infty$.
\end{references}
\end{document}